\documentclass[10pt,preprint2]{aastex}

\shorttitle{}
\shortauthors{Gonz\'alez et al.}

\begin{document}


\title{Gas Dynamical Simulations of the Large and Little Homunculus Nebulae of 
$\eta$ Carinae}


\author{R.F. Gonz\'alez\altaffilmark{1}, E.M. de Gouveia Dal
Pino\altaffilmark{1}, A.C. Raga\altaffilmark{2} 
and P.F. Velazquez\altaffilmark{2}}

\altaffiltext{1}{Instituto Astron\^omico e Geof\'{\i}sico (USP),\\
R. do Mat\~ao 1226, 05508-090 S\~ao Paulo, SP, Brasil\\
e-mail: rgonzalez@astro.iag.usp.br, dalpino@astro.iag.usp.br}

\altaffiltext{2}{Instituto de Ciencias Nucleares (UNAM),\\
Ap. Postal 70-543, CP:04510, M\'exico D.F., M\'exico\\
e-mail: raga@nuclecu.unam.mx, pablo@nuclecu.unam.mx}



\begin{abstract}

We here present two-dimensional, time-dependent radiatively cooling 
hydrodynamical simulations of the large and little Homunculus
nebulae around $\eta$ Carinae. We employ an alternative scenario to previous 
interacting stellar wind models which is supported by both theoretical and 
observational evidence, where a non-spherical outburst wind
(with a latitudinal velocity dependence that matches the observations
of the large Homunculus), which is expelled for 20 years, interacts with
a pre-eruptive slow wind also with a toroidal density distribution, but
with a much smaller equator-to-polar density contrast than that assumed
in previous models. A second eruptive wind with spherical shape is ejected
about 50 years after the first outburst, and causes the development of the
little internal nebula. We find that, as a result of an appropriate 
combination of the parameters that control the degree of asymmetry of the 
interacting winds, we are able to produce not only the structure and
kinematics of both Homunculus, but also the high-velocity equatorial ejecta.
These arise from the impact between the non-spherical outburst and  the
pre-outburst winds in the equatorial plane.

\end{abstract}




\section{Introduction}

The dusty large ($16"$ long) bipolar Homunculus ejecta around $\eta$ Car
is a hollow reflection nebula (e.g., Smith et al. 2003a and references therein)
that was produced by the 20 year Great Eruption of the star, from  $\sim$1840
to 1860. Its axis is inclined  $\sim 45^{o}$ to the line of sight (Davidson et
al. 2001), and at a distance of $\sim$2.3 kpc, it has a total physical size
$\sim 6\times 10^{17}$ cm. The hot  features observed outside the bipolar
nebula were very probably ejected earlier in episodic events before the major
eruption (Walborn et al. 1978; Weis et al. 2001). Recently, an inner 
bipolar emission nebula
 has been
also discovered embedded within the larger
Homunculus (extending from $-2"$ to $+2"$ across the star) which may
have been originated from a minor eruption event in the 1890s, i.e., $\sim$50
years after the formation of the larger Homunculus (Ishibashi et al. 2003).
This little Homunculus seems to follow approximately the shape of the 
larger one. Observations also evidence the existence of ejecta in the equatorial
region that may contain material from both the 1890 eruption and the great 
eruption in the 1840s (Davidson et al. 2001). The velocity of this material 
obtained from H$\alpha$ profiles may reach velocities $\sim 400-750$ km s$^{-1}$ 
(Smith et al. 2003a).

As an extreme luminous blue variable star (LBV), $\eta$ Car loses copious 
amounts of mass
in form of quasi-steady winds punctuated by eruptive events where the mass loss
may increase by at least an order of magnitude in short periods of time
(Maeder 1989; Pasquali et al. 1997). Currently, the dominant velocities in
the expanding Homunculus are 400 to 600 km s$^{-1}$, but some faster material
($\sim$ 1000 km s$^{-1}$) has been detected in the poles (e.g., Smith et al. 
2003a), and the mass-loss rate is $\sim 10^{-4}-10^{-3}$ M$_{\odot}$ yr$^{-1}$ 
(Humphreys \& Davidson 1994, Hillier et al. 2001; Corcoran et al. 2001; Soker 
2001). The 
inner bipolar nebula, produced during the 1890s eruption, has a peak velocity 
$\sim$ 
300 km s$^{-1}$ and a total estimated mass $\sim$ 0.1 M$_{\odot}$ (Ishibashi
et al. 2003). 

It has been previously suggested that the shaping of the $\eta$ Car nebulae could be explained by a colliding wind 
binary star model. Soker (2001), for example, has argued that the companion star could  divert the wind blown by the 
primary star, by accreting from the wind and  by blowing its own collimated fast wind that could have, in turn, played a 
role in the formation of the Homunculus lobes.
A strong argument against a companion star dominating the wind structure is that it appears to be predominantly 
symmetric, as indicated  by recent STIS spectral 
observations in several positions along the Homunculus  
(Smith et al. 2003a).  These  show  the same 
latitudinal dependence for the velocities in both hemispheres and both sides of 
the polar axis, and the same P Cygni absorption in hydrogen lines on either side 
of the poles.
Although a colliding wind binary model cannot be disregarded at the present,  we here assume that the shaping of  
$\eta$ 
Car nebulae is dominated by the primary star's wind.

As noticed by Smith et al. (2003a), the high velocities seen in reflected light 
from the polar lobes give a first direct evidence that the polar axis of the 
Homunculus is aligned with the rotation axis of the central star. This has 
important consequences for the formation of the bipolar lobes and the equatorial 
ejecta around $\eta$ Car, as it may be an indication that axial symmetry and the 
ejection mechanism during the Great Eruption were directly linked to the central 
star's rotation. Also, the observed latitudinal variations in H and HeI lines 
revealing that the speed, density and ionization in $\eta$ Car wind are
non-spherical nearby the star, may be an indication that the stellar wind
is inherently non-spherical.

In previous work, Frank et al. (1995) have performed low resolution 
two-dimensional numerical simulations of the large Homunculus of $\eta$ Car,
adopting an interacting stellar wind scenario wherein a spherical fast
wind expands into a non-spherical (toroidal) slow, dense wind previously
ejected from the star. They found that the Homunculus morphology could be
reproduced with an equator-to-polar density ratio $\sim 200$, which would
imply the existence of a very dense toroidal environment surrounding the
nebula. Langer et al. (1999), have assumed a variant of this scenario
including the effects of stellar rotation. Using the wind-compressed
model of Bjorkman \& Cassinelli (1992), they showed that a strong
equator-to-pole density contrast could have formed during the great outburst
in the 1840s if the star was close to the Eddington luminosity limit. Under
this circumstance, the centrifugal and radiative forces must balance gravity
at the equator, and a strong non-spherical mass loss should occur deflecting
the wind streamlines towards the equator. In this model, a spherical, fast
post-outburst wind produces the bipolar bubble through interaction
with the toroidal outburst flow. In addition to forming lobes with an 
approximate shape to the observed Homunculus, this numerical model (which has
included the effects of time-dependent radiative cooling) has revealed the
development of small fingers at the shell surface caused by Vishniac type
instabilities. 

Although these previous models are partially successful at reproducing the
basic shape of the large Homunculus nebula, they both rely on the 
presence of a thick torus around the Homunculus nebula (which is very dense in
the equatorial region). Observations however, indicate only the presence of a 
faint nebulosity surrounding the large Homunculus. To overcome this difficulty,
more recently Frank et al. (1998), and Dwarkadas \& Balick (1999) have proposed
alternative models. Assuming an inverted scenario, in which a $non-spherical$
fast wind expands into a previously deposited $isotropic$ slow wind, Frank et 
al. (1998) have found that they are able to reproduce strongly bipolar outflows 
with
polar caps that can be denser than the lobes' flanks. However, like the previous
ones, this model is unable to produce the equatorial ejecta. Dwarkadas \& Balick 
(1998), on the other hand, have replaced the thick torus of the previous models 
by a
small and dense, near-nuclear toroidal ring. This also manages to provide some
collimation of the spherical wind ejected during the Great Eruption. Besides,
in the presence of radiative cooling, the ring is completely destroyed by the
impact of the wind, and the authors have claimed that this fragmentation of the
ring could help to explain the equatorial ejecta. A potential problem with this
interpretation is that their model predicts velocities for the fragments that
are too small ($\sim 50 - 100$ km s$^{-1}$) compared to the observed ones in the
outer parts of the ejecta (e.g., Smith et al. 2003a).   

A successful modeling of the formation of the large and little Homunculus 
nebulae around $\eta$ Car should account for both the bipolar morphology and the 
equatorial ejecta. We here present results of numerical simulations that 
consider
an alternative scenario to the interacting stellar winds models above, in
which a fast, non-spherical wind is ejected for 20 years (with a latitudinal
velocity dependence that matches the observations in the large Homunculus),
interacts with a pre-outburst slow wind also with a toroidal density 
distribution,
but with a much smaller equator-to-polar density contrast than that assumed in
previous models. A second eruptive wind with spherical shape is ejected about
50 years after the first outburst, and causes the development of the little
internal nebula. We find that, as a result of an appropriate combination of
the parameters that control the degree of asymmetry of the interacting winds,
this model is able to produce not only the structure of both Homunculus nebulae, 
but
also the equatorial ejecta (see below).
\footnote{We note that there is a number  of proposed mechanisms
in the literature that predict the development of intrinsically non-spherical
winds coming out from rotating LBV stars (see, e.g., Bjorkman \& Cassinelli 
1992; 
Lamers \& Pauldrach 1991; Owocki et al. 1996, 1998).}

We have carried out several hydrodynamical simulations considering various 
possible scenarios for the degree of asymmetry of the interacting winds, but
here, we will present only the model that has best matched the observations.
A more detailed description of the other models
will be presented in a forthcoming paper (Gonzalez, de Gouveia
Dal Pino, Raga \& Velazquez 2003). In contrast to the previous works,
our simulations compute explicitly the time dependent radiative cooling
of the gas including several atomic and ionic species, which allow for a more
realistic evaluation of its effects on the flow.

\section{Numerical Method, Initial Physical Conditions, and Results}

In order to investigate the winds' interaction, we have performed gasdynamic 
simulations using a modified two-dimensional version of the  Yguaz\'u-a adaptive
grid code originally developed by Raga et al. (2000; see also Raga et al. 2002,
Masciadri et al. 2002, Vel\'azquez et al. 2003). This code
integrates the hydrodynamic equations explicitly accounting for the radiative
cooling together with a set of continuity equations for several atomic/ionic
species employing the flux-vector splitting algorithm of Van Leer (1982).
The following species have been considered: H I, H II,
He I, He II, He III, C II, C III, C IV, O I, O II, and O III.
The calculations were performed on a five-level, binary adaptive grid with a 
maximum resolution along the x and y axes of $7.81 \times 10^{14}$ cm. The
computational domain extends over ($4 \times 10^{17}) \times (4 \times 10^{17})$
cm, corresponding to $512 \times 512$ grid points at the highest resolution
grid level.


We assume that a light, hot gaseous toroidal distribution was formed around
$\eta$ Car prior to the Great Eruption of $\sim$1840 (e.g., Weis 2001). To
produce it, a slow and steady wind is assumed to emanate from the stellar
surface into the ambient medium (with initial temperature $T_a=100$ K and
number density $n_a=10^{-3}$cm$^{-3}$), with the following properties
 (see Frank et al. 1995),

\begin{equation}
n= n_0  \biggl({{r_0}\over{r}} \bigg)^2
		 {{1}\over{F_{\theta}}},
\label{1}
\end{equation}

\begin{equation}
v= v_0 F_{\theta},
\label{2}
\end{equation}

\noindent
where $r_0$ is the injection radius, $v_0$ and $n_0$ are the velocity and the
number density at the pole, respectively.
We have considered a similar
function to that adopted by Frank et al. (1995, 1998) to produce a smooth
variation in the wind density and velocity from the equator to the pole,
$F_{\theta}=1-\alpha[(1-e^{-2\beta sin^2\theta})/(1-e^{-2\beta})]$, where
$\theta$ is the polar angle, the parameter $\beta$ controls the shape of the
wind and $\alpha$ the pole-to-equator density and velocity contrasts. The
density and velocity angular dependence in equations (\ref{1}) and (\ref{2}),
respectively, imply a constant mass-loss rate ($\dot M \propto n\; v$) as a 
function
of the angle. 

We have assumed for the pre-eruptive wind a mass
loss rate $\dot M = 10^{-3} \; M_{\odot}$ yr$^{-1}$; $v_{0} = 250$ km s$^{-1}$;
$\alpha = 0.9$; $\beta = 1.5$; $r_0 = 1 \times 10^{16}$ cm; and initial
temperature $T_0 = 10^4$ K. When the slow wind reaches the edge of the
computational domain in the polar direction (y-axis), a non-spherical outburst
wind is turned on, for 20 years, to produce the Great Eruption, with
$\dot M \simeq 7\times 10^{-2} M_{\odot}$ yr$^{-1}$; $v_{0} = 715$ km s$^{-1}$;
and $T_0 = 10^ 4$ K. For simplicity, we have adopted the same functional
dependence [$F(\theta)$] as above to compute the velocity and density angular
variations in  this wind, but with different values for $\alpha$ and $\beta$
(0.78 and 0.3, respectively). These values were obtained from the best fit
of equation (\ref{2}) to the observed latitudinal variation of the expansion
velocity of the large Homunculus (Smith 2002; Davidson et al. 2001).
After 20 years, a third wind with the same conditions of the 
original slow wind 
resumes for 30 years and then, another outburst (which was assumed, for 
simplicity, to be spherical) is allowed to occur for about 10 years with
$\dot M = 10^{-2} \, M_{\odot}$ yr$^{-1}$; and $v_{0} = 317 $ km s$^{-1}$,
after which the original slow wind again resumes. 
The adopted values of $\alpha$ and $\beta$ for the slow pre-outburst wind  
imply a much smaller (larger) equator-to-pole density (velocity) contrast
(= 10) than that used by Frank et al. (1995). This results a fainter and
lighter toroidal envelope around the large Homunculus, as required
by the observations. 


Figures \ref{f1} and \ref{f2} depict the results of this simulation with the
interaction of the five winds above. We notice that two bipolar expanding
shells develop both with shapes and kinematics similar to the large and the
little Homunculus. As in previous calculations (e.g., Frank et al. 1995, 1998),
we find that during the 160 years of evolution, the momentum flux is dominated
by the outbursts, so that the post-outburst slow winds have no significant
effect on the formation of the Homunculus structures. In the case of the
outer Homunculus, since its mass-loss rate is almost an order 
of magnitude larger than that of the pre-outburst wind, it initially expands 
almost ballistically (in the polar direction), without being much decelerated 
by the pre-outburst slow wind, and retains most of the non-sphericity imprinted 
in it near the star, as observed. In the equatorial direction, this first 
outburst wind reaches the pre-eruptive toroid when its shock front is still 
within the computational domain. The impact of the two fronts (at $t\simeq 100$ 
yr after the great eruption; see Fig. 1c) causes the formation of a faint 
equatorial ejection that could explain the outer parts of the observed
equatorial skirt in $\eta$ Car (Fig. 1d). This ejection moves with a mean
velocity $\simeq$ 700 km s$^{-1}$, therefore in qualitative agreement with
the observations (Smith et al. 2003a).     

The internal bubble that develops from the impact of the second (spherical) 
outburst with a pre-outburst wind ejected with the same characteristics of the 
first slow wind (Fig. 1b), soon
acquires a bipolar shape (Fig. 1c) which is very similar to that
implied from the observations of the little Homunculus with an expansion 
velocity $\sim 300$ km s$^{-1}$ (Ishibashi et al. 2003). 
 
The analysis of the results at the outer Homunculus shows that
a double-shock structure has developed  with an outward moving shock 
that sweeps the material of the precursor wind and an inward shock that
decelerates the outburst wind material coming behind. A thin dense and
cold shell develops at the edge of the nebula as a result of the strong
compression that follows the radiative cooling of the shocked
material behind both shocks. For a strong shock (with $v_s\ > 80$ km s$^{-1}$),
the radiative cooling time for an one-dimensional shock is given by 
$t_c \simeq 320$yr $\, v_{s,100}^{1.12}\,\rho_{pre,-22}^{-1}$, where
$v_{s,100}$ is the shock speed in units of 100 km s$^{-1}$ and $\rho_{pre}$ is 
the pre-shock densitiy in units of $10^{-22}$ g cm$^{-3}$ (Hartigan, Raymond
$\&$ Hartmann 1987; Gonzalez 2002). At the poles, the inward shock speed in the
simulation is   $v_c \simeq 80$ km s$^{-1}$, and $\rho_{pre} \simeq 5.1 \times
10^{-20}$ g cm$^{-3}$. This implies $t_c \simeq 0.5$ yr, which is very short
compared to the age of the large Homunculus bubble ($\sim 160$ yr), and results
in very small cooling distance behind the inner shock that can be estimated from 
$d_c\simeq  3.7 \times 10^{14}$ cm $ v_{s,100}^{4.73}\, \rho_{pre,-22}^{-1}$
(Hartigan, Raymond $\&$ Hartmann 1987), or $d_c\simeq  2 \times 10^{11}$ cm
behind the inward shock. This value qualitatively explains the narrowness of
the cold thin shell seen in the simulations (Fig. \ref{f1}), which, at $t= 160$ 
yr,
has a temperature $\sim 6000$ K and a density $3 \times 10^{-19}$ g cm$^{-3}$. 
Behind the outward  shock, on the other hand, the smaller pre-shock density  
($\rho_{pre} \simeq 2.1 \times 10^{-21}$ g cm$^{-3}$) and the higher shock
speed ($v_c \simeq 400$ km s$^{-1}$) produce a larger cooling time
($t_c \simeq 69$ yr) and, as a result, a thicker and hotter polar cap
with $d_c \simeq 10^{16}$ cm (according to the equation above), and
$ T\simeq 1.6\times 10^4$ K and  $\rho\simeq$ $2.1 \times 10^{-21}$g cm$^{-3}$
(as obtained from the simulations in Fig. \ref{f1}). Despite the approximations 
involved
in the evaluation of the cooling distance above, it is only a factor three 
smaller 
than the thickness of the outer shell obtained from the simulations which is 
$\sim 3.5 \times 10^{16}$ cm. This value is in turn, comparable to the 
observations
that indicate a radial thickness of the outer Homunculus $\sim 1 '' = 2500$ AU
at the poles (Smith et al. 2003a).

A similar analysis for the expanding shell of the inner Homunculus shows that 
its  thickness at the poles (obtained from the simulations) is 
$\sim$10$^{16}$ cm.

\section{Discussion and Conclusions}

The results of our 2-D hydrodynamical simulations involving the interaction of 
five winds with different initial conditions indicate that the shape and 
kinematics of the large Homunculus of $\eta$ Car can result from 
the interaction between fast and slow intrinsically $non-spherical$ winds. This  
model is a variant of previous interacting wind scenarios that have assumed 
either fast spherical winds interacting with a dense and heavy toroidal 
environment (Frank et al. 1995; Langer et al. 1999; Dwarkadas \& Balick 1999) 
or fast non-spherical winds interacting with an isotropic environment (Frank et 
al. 1998). It has two attractive advantages: (i) it shows that
a non-spherical fast wind impinging on a slow toroidal wind 
is able to produce the high-velocity outer parts of the equatorial ejecta 
observed around $\eta$ Car (Smith et al. 2003a); and (ii) the choice of a
lighter pre-outburst wind in our model, has resulted in a less dense toroidal
halo around the large Homunculus nebula than in previous models, as required by 
the
observations.
\footnote{ We note that the equatorial waist produced in the simulations 
is slightly thicker than the one observed. While observations indicate that the 
ejections are $<$0.4 times the diameter of the Homunculus lobes, 
in the simulations this is $\sim$0.6. We also note that 
the ratio of the length to the width of the two Homunculus is somewhat larger 
in the simulations. Both differences 
could be partially attributed to 
projection effects which were not considered here, and 
could be diminished with 
minor changes in the adopted  parameters for the model.}

It is noteworthy that, in numerical experiments where, instead of an initially 
non-spherical, a spherical outburst wind was injected into a toroidal 
pre-outburst slow wind with the asymmetry parameters $\alpha$ and $\beta$ 
determined
either from the observed large Homunculus expansion velocity distribution
or given by the same values as those used by Frank et al. (1995) for a heavier
toroid, have failed to produce simultaneously the shape of the large
Homunculus and the equatorial ejecta. In these cases, the encounter of the  
shock fronts of the two winds first in the equatorial region and afterwards in
the lateral regions of the outer bubble causes its fragmentation and spreading 
of
the shell material at high latitudes, thus destroying the bipolar morphology
(see Gonzalez et al. 2003). These results suggest that, in order to produce
both the Homunculus bipolar morphology and the equatorial ejection from the
winds interaction, these must be both intrinsically non-spherical (as in
Figs. \ref{f1} and \ref{f2}). 

Although the interaction of the second outburst wind (assumed to be spherical) 
with its pre-outburst wind was able to produce the internal Homunculus, it has 
failed to develop internal equatorial ejecta in the simulation depicted in 
Figures \ref{f1} and \ref{f2}. However, an appropriate combination of
non-spherical wind parameters in this case similar to the one of the 1840s
outburst, could also probably generate an internal equatorial ejection. In fact,
recent UV images within 0.2 arcsec of the star, have revealed the existence
of a little internal torus that may be related to the little Homunculus and
may signify that a recurrent mass ejection with the same geometry as that of
the Great Eruption may have occurred (Smith et al. 2003b, see also Gonzalez
et al. 2003).

In order to simulate an experiment using a condition at the base of the $\eta$ 
Car wind similar to the one $presently$ suggested by the observations (Smith 
et al. 2003a),  we have also computed a model in which the non-spherical 
outburst wind of 1840s impinges on a slow pre-outburst wind with a larger 
density 
(and mass-loss rate) in the polar direction ($n\propto F_{\theta}$
in eq. [\ref{1}]). We find that this scenario is unable to develop a narrow 
equatorial ejecta. This result suggests that the conditions at the wind base
prior to the Great Eruption in the 1840s were probably not the same as the
current ones.

Finally, we notice that, despite the high-resolution and the explicit 
time-dependent computation of the radiative cooling of the gas, our simulations, 
similarly to the radiative cooling models of Frank et al. (1998), have $not$ 
revealed the formation of small fragments on the surface of the Homunculus,
as those seen in Langer et al. (1999) simulations, which have resulted
from Vishniac instabilities in the radiatively cooled shell
long after the eruption.
This is probably due to differences in the initial conditions between the two 
models. Nonetheless, a granular structure is effectively observed on the 
large Homunculus surface. Is is not improbable, however, that 
they have resulted from variability or instabilities in the winds $near$ the
surface of the star (Smith et al. 2003a). This question, as well as 
three-dimensional effects,
will be addressed in future work.

\acknowledgments{This work has been partially supported by grants of the 
Brazilian Agencies FAPESP and CNPq, and by the
Mexican CONACyT fellowship 020179,  the 
CONACyT 
grants
36572-E and 41320, and the DGAPA (UNAM) grant IN112602. 
The authors have
benefited from elucidating conversations and comments from Augusto Damineli, 
Nathan Smith, and the referee Noam Soker.}

\clearpage

\begin{figure}
\caption{\small  Gray-scale map of the density distribution (in $log_{10}$ 
scale) for four different times in the evolution
of the model. (a): t = 10 yr; (b) t = 60 yr; (c) t = 110 yr; and (d) t = 160 yr.
The density (at the vertical scale on the right side of the figures) is in g 
cm$^{-3}$, and the x and y axis are in cm.
\label{f1}}
\end{figure}

\begin{figure}
\caption{\small The same as in Figure 1d where it is depicted the present-day
structure of the system with the inner and outer Homunculus nebulae, and the
equatorial ejections. The density map has been rotated by an angle of 45$^{\sf 
o}$,
like in the observations.
\label{f2}}
\end{figure}


\begin{thebibliography}{}
\small
\bibitem[]{484}
Bjorkman, J.E., $\&$ Cassinelli, J.P. 1992, ASP Conference Series,
Vol. 22, 88
\bibitem[]{491}
Corcoran, M.F., Ishibashi, K., Swank, J.H., $\&$ Petre, R. 2001,
\apj, 547, 1034
\bibitem[]{498}
Davidson, K., Smith, N., Gull, T.R, Ishibashi, K., $\&$ Hillier, D.J. 2001,
\aj, 121, 1569
\bibitem[]{501}
Dwarkadas, V.V., $\&$ Balick, B. 1998, \aj, 116, 829 
\bibitem[]{503}
Frank, A., Balick, B., $\&$ Davidson, K. 1995, \apjl, 441, L77
\bibitem[]{505}
Frank, A., Ryu, D., Davidson, K. 1998, \apj, 500, 291
\bibitem[]{507}
Gonz\'alez, R.F. 2002, Ph.D. thesis, Univ. Nacional Aut\'onoma de M\'exico
\bibitem[]{509}
Gonz\'alez, R.F., de Gouveia Dal Pino, E.M., Raga, A.C., 
$\&$ Vel\'azquez, P.F. 2003, (in prep.)
\bibitem[]{512}
Hartigan, P., Raymond, J., \& Hartmann, L. 1987, \apj, 316, 323
\bibitem[]{516}
Hillier, D.J., Davidson, K., Ishibashi, K., $\&$ Gull, T. 2001, \apj, 553, 837
\bibitem[]{518}
Humphreys, R.M., $\&$ Davidson, K. 1994, PASP, 106, 1025
\bibitem[]{520}
Ishibashi, K., et al. 
2003, \aj, 125, 3222 
\bibitem[]{527}
Lamers, H.J.G.L.M., $\&$ Pauldrach, A.W.A. 1991, A$\&$A, 244, L5  
\bibitem[]{529}
Langer, N., Garc\'ia-Segura, G., $\&$ Mac Low, M. 1999, \apjl, 520, L49
\bibitem[]{531}
Masciadri, E., de Gouveia Dal Pino, E.M., Raga, A.C., $\&$ Noriega-Crespo, A.
2002, \apj, 580, 950
\bibitem[]{534}
Maeder, A. 1989, in The Physics of Luminous Blue Variables, ed. K.
Davidson, A.F.J. Moffat, $\&$ H.J.G.L.M. Lamers (Dordrecht:Kluwer), 15  
\bibitem[]{537}
Owocki, S.P., Cranmer, S.R., $\&$ Gayley, K.G. 1996, \apj, 472, L115 
\bibitem[]{539}
Owocki, S.P., Gayley, K.G., $\&$ Cranmer, S.R. 1998, in ASP Conf. Ser. 131,
Boulder Munich II: Properties of Hot Luminous Stars, ed. I.D. Howarth
(San Francisco: ASP), 237
\bibitem[]{543}
Pasquali, A., Langer, N., Schmutz, W., Leitherer, C., Nota, A., Hubeny, I.,
$\&$ Moffat, A.F.J. 1997, \apj, 478, 340
\bibitem[]{546}
Raga, A.C., de Gouveia Dal Pino, E.M., Noriega-Crespo, A., Mininni, P.D.,
$\&$ Vel\'azquez, P.F. 2002, A$\&$A, 392, 267
\bibitem[]{549}
Raga, A.C, Navarro-Gonz\'alez, R., $\&$ Vilagr\'an-Muniz, M.
2000, Rev. Mex. Astron. Astrof., 36, 67
\bibitem[]{554}
Smith, N. 2002, \mnras, 337,1252
\bibitem[]{556}
Smith, N., Davidson, K., Gull, T.R., Ishibashi, K., $\&$ Hillier, D.J.
2003a, \apj, 586,432
\bibitem[]{559}
Smith, N., Morse, J.A., Gull, T.R., Hillier, J. et al. 2003b, \apj, 
(submitted)
\bibitem[]{562}
Soker, N. 2001, \mnras, 325, 584
\bibitem[]{564}
Van Leer, B. 1982, ICASE Report No. 82-30
\bibitem[]{566}
Vel\'azquez P.F., Koenigsberger G., $\&$ Raga A.C. 2003, \apj, 584, 284
\bibitem[]{568}
Walborn, 1978, \apj, 219, 498
\bibitem[]{566}
Weis, K., Duschl, W.J., $\&$ Bomans, D.J. 2001, A$\&$A, 367, 566


\end{thebibliography}
\end{document}